\documentclass[sigconf,nonacm]{acmart}

\usepackage{enumitem}

\hypersetup{keeppdfinfo}
\newcommand{\secref}[1]{Section~\ref{#1}}

\def\system{MultiMoQ}

\title{\system{}: Multi-Access Media-Over-QUIC for Robust Immersive Video Streaming}

\setcopyright{cc}
\setcctype{by}

\author{Yitong Li}
\orcid{0009-0008-4728-449X}
\affiliation{%
  \institution{The Hong Kong University of Science and Technology (Guangzhou)}
  \city{Guangzhou}
  \state{Guangdong}
  \country{China}
}
\email{yliop@connect.hkust-gz.edu.cn}

\author{Xinjiao Li}
\orcid{0009-0003-9591-4031}
\affiliation{%
  \institution{The Hong Kong University of Science and Technology (Guangzhou)}
  \city{Guangzhou}
  \state{Guangdong}
  \country{China}
}
\email{xli886@connect.hkust-gz.edu.cn}

\author{Ruonan Chai}
\orcid{0000-0003-1821-0914}
\affiliation{%
  \institution{The Hong Kong University of Science and Technology (Guangzhou)}
  \city{Guangzhou}
  \state{Guangdong}
  \country{China}
}
\email{rchai327@connect.hkust-gz.edu.cn}

\author{Dirk Kutscher}
\correspondingauthor
\orcid{0000-0002-9021-9916}
\affiliation{%
  \institution{The Hong Kong University of Science and Technology (Guangzhou)}
  \city{Guangzhou}
  \state{Guangdong}
  \country{China}
}
\email{dku@hkust-gz.edu.cn}

\renewcommand{\shortauthors}{Li et al.}



\begin{document}

\begin{abstract}

Live immersive video streaming, particularly 360\textdegree{} video, is increasingly adopted in applications such as virtual events, sports broadcasting, and remote education.
Existing approaches struggle to support high-bitrate immersive streaming for large numbers of concurrent users, with coarse-grained delivery limiting responsiveness and insufficient support for coordinating concurrent tile streams.
Media over QUIC (MoQ) has recently emerged as a promising solution for large-scale media delivery, yet it lacks robustness under bandwidth-constrained conditions, often resulting in playback stalls.
To address these challenges, we present \system{}, a multi-access tile streaming framework built on MoQ that redesigns its delivery mechanism for robust high-bitrate streaming across multiple access paths while supporting flexible tile scheduling and seamless access switching.
We implement a fully functional prototype of \system{} and evaluate it in network emulation under heterogeneous real-world network conditions, comparing against Dynamic Adaptive Streaming over HTTP (DASH) and standard MoQ.
Results show that \system{} increases goodput for enhancement tiles and base video and reduces enhancement-tile tail end-to-end latency relative to DASH, while preserving audio continuity and avoiding the persistent stalls of standard MoQ.
These transport gains translate into smoother viewport playback, and the ablation results further confirm the contribution of multi-access control to playback continuity.
    
\end{abstract}

\maketitle
\hypersetup{pdfauthor={Yitong Li, Xinjiao Li, Ruonan Chai, Dirk Kutscher}}

\section{Introduction}
\label{sec:intro}

Live 360\textdegree{} video lets viewers explore an omnidirectional scene while it is captured and delivered. It supports concerts, sports broadcasts, virtual exhibitions, and remote education~\cite{360video-survey1,360video-survey2,360video-cultural,360video-livesports,360video-education}. Unlike conventional video, it carries a high-resolution panoramic scene while each viewer selects a changing viewport. Live operation therefore requires timely and continuous delivery so that viewport changes and playback remain responsive under time-varying network conditions~\cite{360video-survey2}.

Prior work addresses parts of this problem. Viewport-adaptive systems reduce traffic by sending only tiles relevant to the current or predicted view, but commonly use DASH-style HTTP delivery~\cite{taxonomy-applayer1,taxonomy-applayer2,taxonomy-applayer3,taxonomy-applayer4}. QUIC and multipath-QUIC designs improve transport efficiency and path utilization~\cite{taxonomy-transport1,taxonomy-transport2,taxonomy-transport3}. Multi-source, multicast, and edge-assisted systems distribute load across network resources~\cite{taxonomy-other1,taxonomy-other3,taxonomy-other-edge1}. However, these approaches do not jointly provide responsive live delivery, scalable fan-out, and robust coordination of concurrent tile streams under heterogeneous access conditions.

Three challenges remain. First, low-latency playback requires finer-grained delivery than conventional segment mechanisms provide, especially as access quality changes. Second, direct client--server retrieval creates repeated upstream traffic when many users request high-bitrate content. Third, playback depends on several concurrent tile streams; instability in even a subset can create missing regions or freeze the reconstructed viewport. A viable design must address latency, robustness, and scalability together rather than optimizing one stream or path in isolation.

\begin{figure}[t]
  \centering
  \includegraphics[width=1\linewidth]{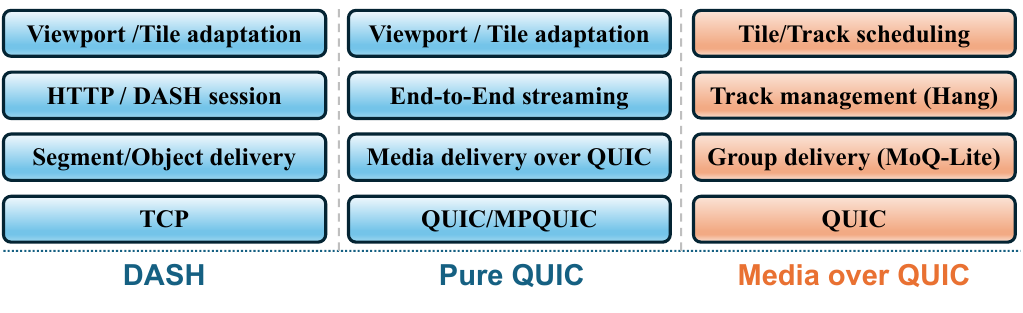}
  \caption{High-level comparison of DASH-, QUIC-, and MoQ-based immersive streaming architectures.}
  \Description{Three side-by-side protocol stacks compare immersive streaming architectures. The DASH stack contains viewport or tile adaptation, an HTTP/DASH session, segment or object delivery, and TCP. The pure-QUIC stack contains viewport or tile adaptation, end-to-end streaming, media delivery over QUIC, and QUIC or multipath QUIC. The Media-over-QUIC stack contains tile or track scheduling, Hang track management, MoQ-Lite group delivery, and QUIC.}
  \label{fig:intro_high_level}
\end{figure}

Fig.~\ref{fig:intro_high_level} summarizes the architectural tradeoffs. DASH tile streaming reduces unnecessary media traffic. Its segment requests respond coarsely and can generate repeated backhaul fetches. Pure QUIC designs improve end-to-end transport but retain direct client--server delivery, making fan-out difficult. Network-assisted designs use distributed resources but require stronger support at servers, edges, or the network. Media over QUIC (MoQ)~\cite{moq} instead combines relay-assisted fan-out with object-level delivery. Our measurements in \secref{subsec:motivation}, however, show that standard MoQ remains unstable when it carries many high-bitrate tiles.

We design \system{}, a client-driven multi-access framework built on MoQ. Its first component redesigns MoQ-Lite group delivery with separate pending and active states, preserving in-flight transmission while still discarding stale media. Its second component lets each client monitor delivery quality, warm an alternate access, and switch individual tracks without exposing discontinuities to the application. Together, the components provide continuous multi-track delivery and timely access adaptation while retaining MoQ's relay-assisted architecture.

This division is deliberate. The lower layer handles continuity along the complete server--relay--client path, while the endpoint controller handles changes in access quality. A stable delivery layer is required before switching can improve timeliness without creating another media discontinuity.

This paper makes the following contributions:
\begin{itemize}[leftmargin=*, topsep=2pt, itemsep=1pt, parsep=0pt, partopsep=0pt]
    \item \textbf{Empirical analysis of existing immersive streaming approaches.}
    We identify distinct DASH and standard-MoQ bottlenecks under heterogeneous client conditions and connect them to their delivery architectures (\secref{subsec:motivation}).
    The analysis distinguishes delayed segment retrieval from freshness-driven group interruption.

    \item \textbf{Robust MoQ delivery redesign for concurrent tile streaming.}
    We separate pending-group maintenance from active transmission and schedule groups using freshness and completion gaps (\secref{subsec:lite_redesign}).
    The same queue separation applies at senders and receivers along the relay path.

    \item \textbf{Client-driven multi-access architecture for seamless tile handover.}
    Per-track monitoring and make-before-break multiplexing switch accesses while maintaining one continuous application stream (\secref{subsec:multi_access_control}).
    Clients can retain one access or warm a second according to observed delivery quality.

    \item \textbf{Prototype implementation and evaluation.}
    We extend the Rust MoQ codebase and compare \system{} with DASH and standard MoQ in a 96-client Mininet topology configured from real-world access measurements (\secref{sec:evaluation}).
    Transport measurements, an ablation, and reconstructed playback evaluate the two components from network to application output.
\end{itemize}

\section{Background}
\label{sec:background}

\subsection{360\textdegree{} Video Streaming}
\label{subsec:360video}

An equirectangular projection (ERP) maps a spherical 360\textdegree{} scene to a rectangular video frame~\cite{360video-survey2}. Sending the entire high-resolution ERP frame at uniform quality consumes substantial bandwidth, particularly for live content.

Viewport-adaptive systems instead partition the panorama into spatial tiles. They send tiles in the current or predicted viewport at higher quality and reduce quality elsewhere~\cite{taxonomy-applayer1,taxonomy-applayer2}. This saves bandwidth but turns one video into several media streams whose delivery must remain synchronized as the viewport and network change.

Most deployed designs implement this adaptation through DASH segments~\cite{dash,taxonomy-applayer3,taxonomy-applayer4}. DASH is widely supported, but its segment granularity limits how quickly a live session can react to access variation or a new viewport.

\subsection{Media over QUIC}
\label{subsec:moq}

\begin{figure}[ht]
  \centering
  \includegraphics[width=1\linewidth]{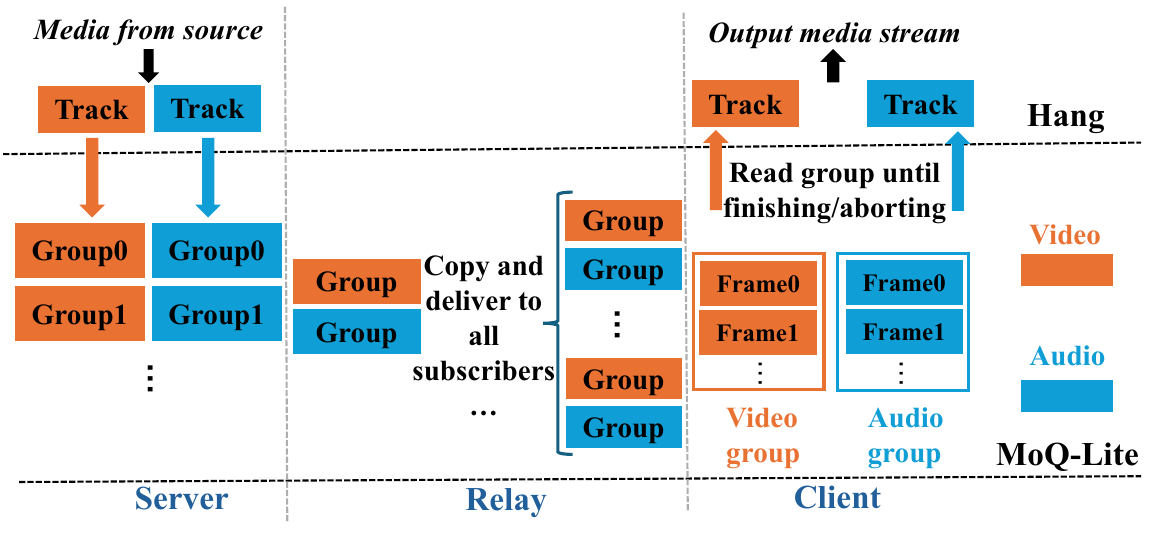}
  \caption{Overview of Media over QUIC.}
  \Description{A left-to-right media-delivery pipeline. A server separates source media into video and audio tracks and groups. A relay copies received groups to all subscribers. At the client, MoQ-Lite receives groups containing frames, while the Hang layer reads each group until completion or abortion and exposes the resulting video and audio tracks as an output media stream.}
  \label{fig:back_overview_moq}
\end{figure}

Media over QUIC (MoQ) is a publish--subscribe framework for real-time media over QUIC or WebTransport~\cite{moq,moq-requirement}. As Fig.~\ref{fig:back_overview_moq} shows, a producer publishes named tracks that intermediate relays distribute to subscribers. Unlike a client issuing an independent HTTP request for every segment, many subscribers can share this relay-assisted distribution tree, providing scalable fan-out.

MoQ organizes each track into groups of media objects~\cite{moq}. A track is a logical media stream, a group is a delivery and scheduling unit, and objects carry media data. This hierarchy supports finer-grained transport than conventional segments and lets different immersive-video tiles use independent tracks.

Independent tracks introduce a continuity requirement. Audio, base video, and enhancement tiles may progress at different rates when bandwidth varies; losing or delaying one track can still disrupt the combined viewport. MoQ must therefore coordinate fresh delivery with completion of in-flight groups.

\subsection{Motivation}
\label{subsec:motivation}

\begin{table}[t]
\centering
\caption{Campus network measurement: bandwidth/delay of Alibaba/Tencent CDN\protect\footnotemark}
\label{tab:campus_network}
\begin{tabular}{c c c c c}
\hline
\textbf{BW Cat.} & \textbf{BW} & \textbf{Prop.} & \textbf{Ali Delay} & \textbf{Tencent Delay} \\
                 & \textbf{(Mbps)} & & \textbf{(ms)} & \textbf{(ms)} \\
\hline
Bad  & 8--15  & 25\% & 11--14 & 15--20 \\
Stable  & 15--30 & 50\% & 10--13  & 15--18 \\
Good & 30--40 & 25\% & 9--13  & 14--18 \\
\hline
\end{tabular}
\end{table}

\footnotetext{Bandwidth and latency measurements were collected across different locations and time periods within the campus network using standard tools such as \texttt{ping} and \texttt{curl}.}

\begin{figure}[ht]
    \centering
    \begin{minipage}[t]{0.52\columnwidth}
        \centering
        \includegraphics[width=0.99\linewidth]{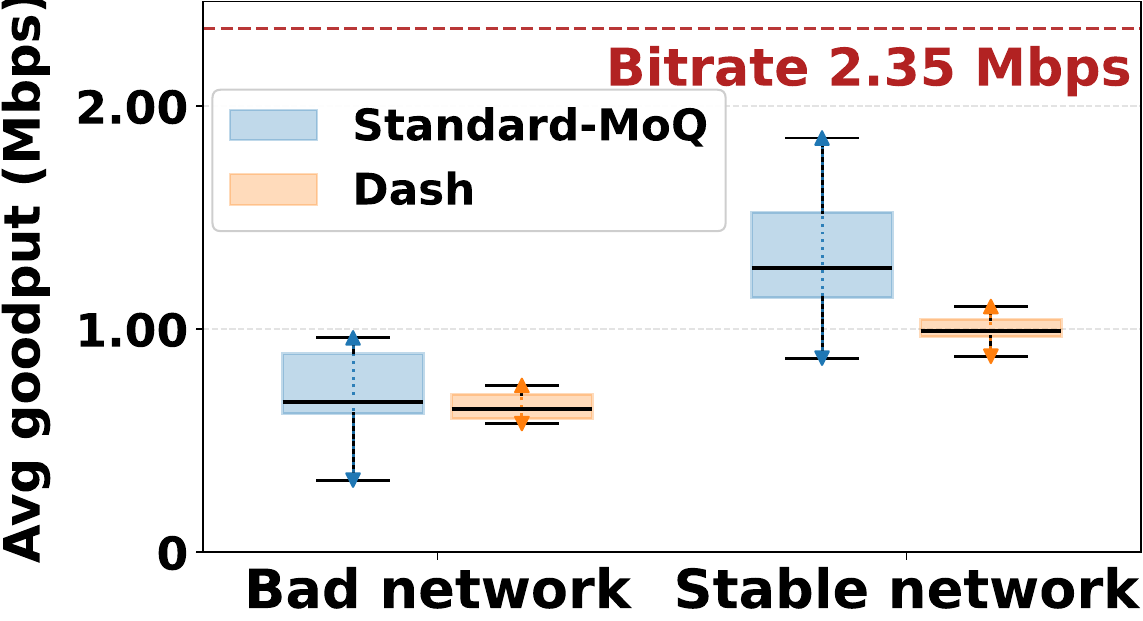}
        \caption{MoQ/DASH goodput.}
        \Description{Two box plots compare per-tile goodput for standard MoQ and DASH under bad and stable access against a 2.35 Mbps target bitrate. Standard MoQ is higher but more variable, and both systems remain below the target.}
        \label{fig:motivation_goodput}
    \end{minipage}%
    \quad
    \begin{minipage}[t]{0.44\columnwidth}
        \centering
        \includegraphics[width=0.99\linewidth]{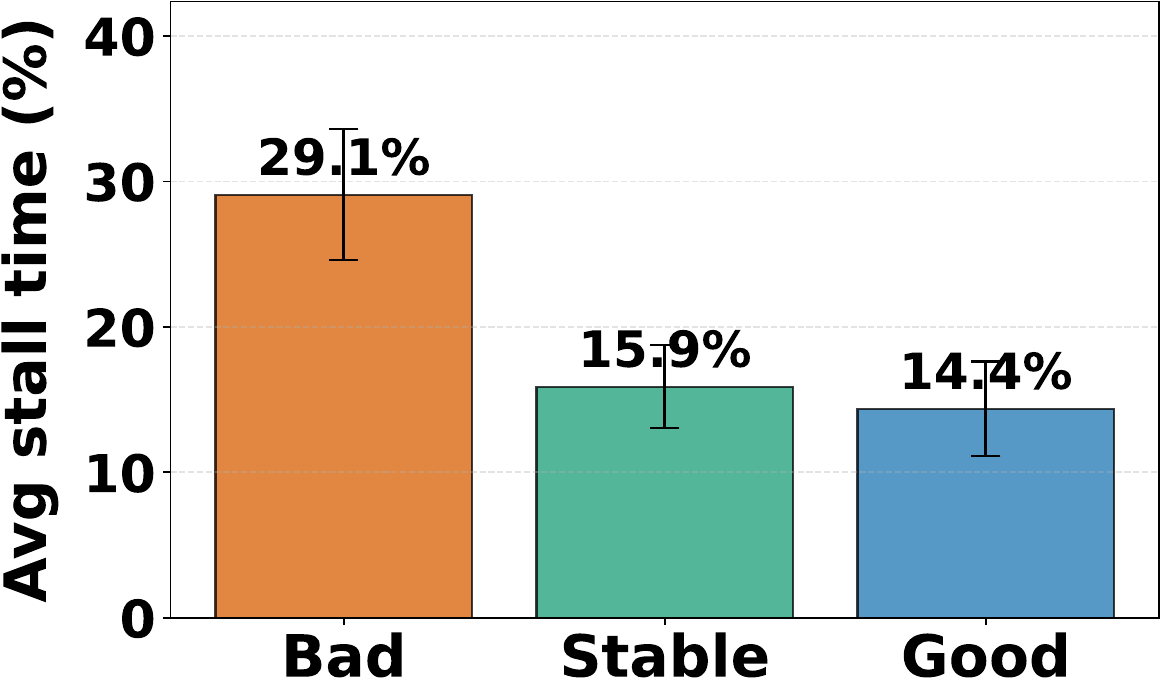} 
        \caption{MoQ stall time.}
        \Description{A bar chart shows standard-MoQ average stall times of 29.1, 15.9, and 14.4 percent under bad, stable, and good access, respectively, with error bars.}
        \label{fig:motivation_stall}
    \end{minipage}
\end{figure}

We first compare DASH and standard MoQ using the measured access conditions in Table~\ref{tab:campus_network}. The experiment serves 96 clients through a four-layer hierarchy of CDN proxies or MoQ relays (\secref{subsubsec:topology}). Both baselines stream the same 4K, 30-fps ERP source, whose enhancement tiles average 2.35~Mbps. DASH retrieves 1-s segments through caching proxies; standard MoQ uses the current IETF architecture and 1-s groups without our modifications. \secref{subsubsec:streaming_setup} gives the complete setup.

DASH goodput remains 20--40\% below the encoding rate in the bad and stable conditions (Fig.~\ref{fig:motivation_goodput}). Standard MoQ approaches the target rate, but its average stall time is 29.1\%, 15.9\%, and 14.4\% in the bad, stable, and good conditions, respectively (Fig.~\ref{fig:motivation_stall}). Thus, DASH loses timeliness and throughput while standard MoQ loses delivery continuity. These different failure modes motivate an architectural rather than bitrate-adaptation change.

\noindent\textbf{\textit{DASH: uncoordinated segment requests create backhaul pressure at scale.}}
DASH clients independently request HTTP segments~\cite{dash}. In a tile-based session, each client requests multiple high-bitrate tracks and each cache miss triggers an upstream fetch. Concurrent misses therefore increase backhaul traffic and compete with downstream delivery. The segment unit also delays reactions until request or segment boundaries. Edge caching and collaborative delivery can reduce remote requests~\cite{taxonomy-other-edge1,taxonomy-other-edge2,taxonomy-other-edge3,taxonomy-other4}, but they do not remove the coarse request granularity~\cite{taxonomy-applayer2,taxonomy-applayer3}. As the client population grows, this model becomes difficult to sustain under a constrained shared cache and backhaul.

\noindent\textbf{\textit{MoQ: freshness-oriented delivery and tight cross-layer coupling hurt robustness.}}
In the current MoQ-Lite design, each track maintains one active group; starting a new group may close or drop an older incomplete group~\cite{moq-lite-rfc}. This policy favors fresh live media, consistent with MoQ's deadline-driven semantics~\cite{moq}. However, replacing an incomplete group can create a gap even when useful objects are still in flight.

The interaction with the Hang layer compounds this behavior. A standard MoQ-Lite receiver exposes one current group and replaces it when a newer group arrives. Hang must consume the current group before replacement, creating a race between lower-layer freshness and upper-layer playback. Short-fragment audio is especially sensitive because groups arrive frequently. With many high-bitrate tile tracks, repeated replacement can interrupt several in-flight groups and propagate instability to the reconstructed viewport. The stalls in Fig.~\ref{fig:motivation_stall} show the resulting loss of continuity.

A robust MoQ design must satisfy two requirements. It must separate freshness maintenance from active transmission so that a new group does not immediately disrupt an in-flight group. It must also adapt away from a degraded access before transport delay becomes playback lag, while presenting each tile as one continuous stream to the application.

These requirements span different layers. Group lifetime and forwarding belong in MoQ-Lite along the relay path, whereas access selection and media cutover require endpoint knowledge of delivery quality and decoder progress. Treating both as one scheduling decision would couple relays to application state; \system{} instead assigns each responsibility to the layer that observes the required signal.

\section{\system{} Design and Implementation}
\label{sec:design}

\begin{figure}[ht]
  \centering
  \includegraphics[width=1\linewidth]{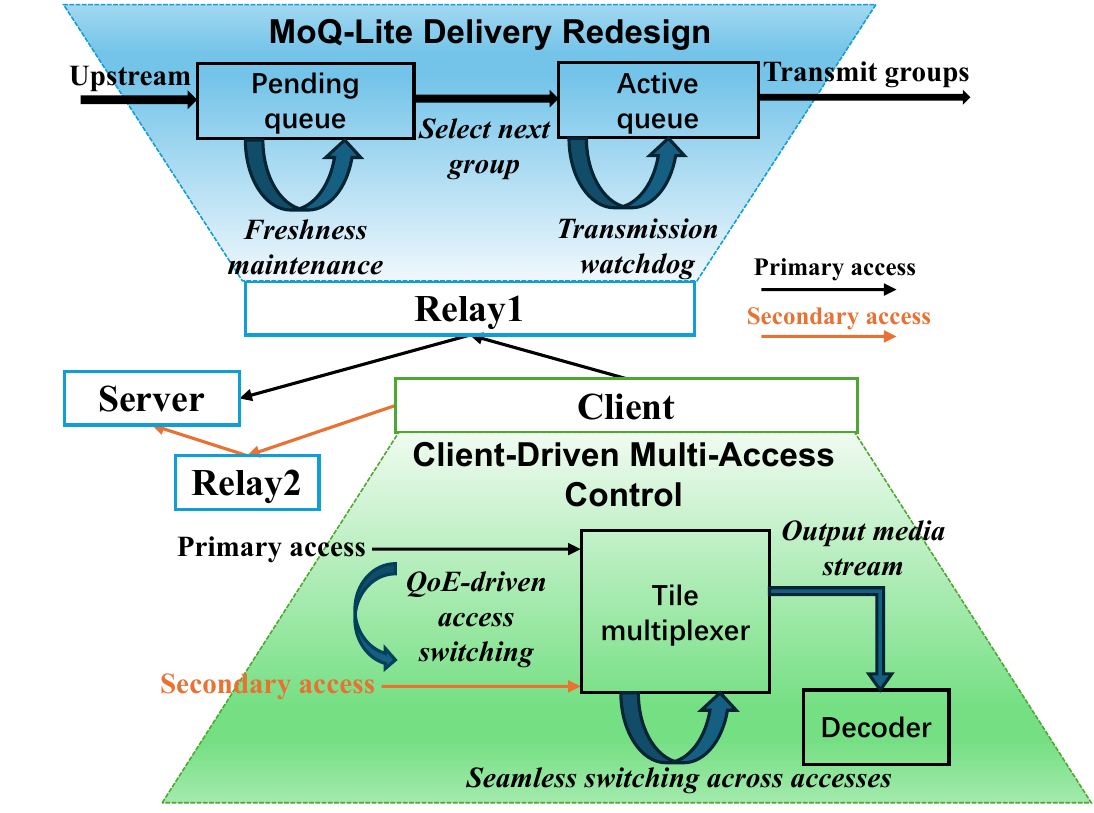}
  \caption{\system{} overview.}
  \Description{MultiMoQ has two connected subsystems. The upper MoQ-Lite delivery redesign passes upstream groups through a freshness-maintained pending queue, selects the next group for an active queue monitored by a transmission watchdog, and transmits groups through a relay. The lower client-driven multi-access subsystem receives black primary-access and orange secondary-access streams, selects them through a quality-driven tile multiplexer, and sends one continuous media stream to the decoder.}
  \label{fig:design_overview}
\end{figure}

\begin{figure*}[!htbp]
    \begin{minipage}[t]{\textwidth}
        \centering
        \includegraphics[width=\textwidth]{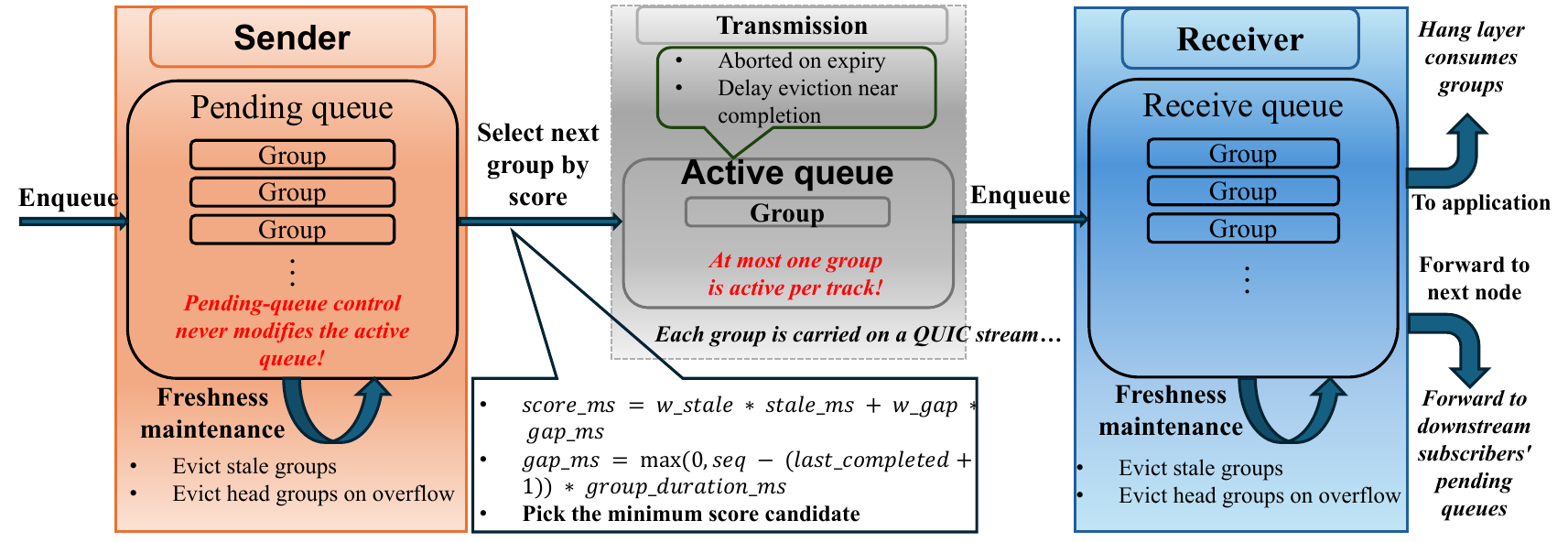}
        \caption{MoQ-Lite delivery redesign.}
        \Description{A sender-to-receiver group-delivery workflow. The sender enqueues groups in a pending queue that evicts stale or overflowed groups without modifying active transmission. A freshness-and-gap score selects the next group, and at most one group per track enters the active queue and its QUIC stream. The receiver enqueues delivered groups in a freshness-maintained receive queue, from which the Hang layer consumes groups locally or a relay forwards them to downstream subscribers.}
        \label{fig:redesign_lite}
    \end{minipage}
\end{figure*}

\subsection{Overview of \system{}}
\label{subsec:design_overview}

Fig.~\ref{fig:design_overview} shows \system{}'s two components. The MoQ-Lite redesign (\secref{subsec:lite_redesign}) separates buffered groups from active transmission and applies this behavior at each sender and receiver along the relay path. Client-driven multi-access control (\secref{subsec:multi_access_control}) builds on the resulting stable delivery at the Hang and application layers. It monitors each track, warms an alternate path when needed, and multiplexes both accesses into one application-visible stream.

\subsection{MoQ-Lite Delivery Redesign}
\label{subsec:lite_redesign}

MoQ-Lite~\cite{moq-lite-rfc} is MoQ's transport-facing layer and the lowest layer that \system{} changes. We modify its group-delivery path to address the freshness--continuity tradeoff in \secref{subsec:motivation}. The redesign preserves MoQ's layered architecture, publish/subscribe semantics, and relay role (\secref{subsec:moq}); it changes how groups wait for and enter transmission.

\paragraph{Sender-side buffering.} As Fig.~\ref{fig:redesign_lite} shows, a FIFO \textit{pending queue} admits newly received groups\footnote{The queue size is tunable; our prototype uses five entries.}. Periodic maintenance removes stale entries, and overflow eviction starts at the queue head. The active transmission state is separate: pending-queue admission, expiration, and overflow handling never modify the \textit{active queue}. A new group can therefore update the sender's candidates without immediately aborting an in-flight group.

\paragraph{Active transmission control.}
The sender scores each pending group to balance freshness against sequential completion. Always selecting the newest group minimizes age, but it can skip unfinished sequence numbers and leave gaps in the client timeline. Always selecting the oldest group preserves order but can deliver obsolete live media. For candidate $g$, Eqs.~\eqref{eq:group_score} and \eqref{eq:group_gap} combine these costs:
\begin{align}
\mathit{score}(g) &= w_{\mathit{stale}} \cdot \mathit{stale}_g
+ w_{\mathit{gap}} \cdot \mathit{gap}_g, \label{eq:group_score}\\
\mathit{gap}_g &= \max\!\left(0,\mathit{seq}_g-(\mathit{last}_{\mathit{comp}}+1)\right)\cdot T_g.
\label{eq:group_gap}
\end{align}
Here, $\mathit{stale}_g$ is staleness in milliseconds, $\mathit{seq}_g$ is the group sequence number, $\mathit{last}_{\mathit{comp}}$ is the last completed sequence number on the track, and $T_g$ is group duration in milliseconds. The gap term expresses the playback time skipped if $g$ follows the last completed group. Experiments use $w_{\mathit{stale}}=1$ and $w_{\mathit{gap}}=0.5$ to favor freshness without ignoring continuity. The sender selects the minimum-score group.

The selected group becomes the track's only active group and uses a dedicated QUIC stream. At most one active group per track keeps group scheduling lightweight and leaves byte transmission to QUIC. Starting many groups in parallel under a constrained path could delay every group and leave only stale partial deliveries. In \system{}, later arrivals remain in the pending queue until the current group completes, expires near its delivery deadline, or stops making progress. The transmission watchdog handles the latter cases and releases the active slot for another candidate.

\paragraph{Receiver delivery.} A freshness-maintained \textit{receive queue} holds delivered groups instead of exposing one replaceable group. At a relay, each received group is copied to the pending queues of downstream subscribers and forwarded to the next node. At a client, the receive queue buffers groups until Hang consumes them. MoQ-Lite remains responsible for transport and forwarding, while media consumption stays in the higher layer.

\subsection{Client-Driven Multi-Access Control}
\label{subsec:multi_access_control}

\begin{figure}[ht]
  \centering
  \includegraphics[width=1\linewidth]{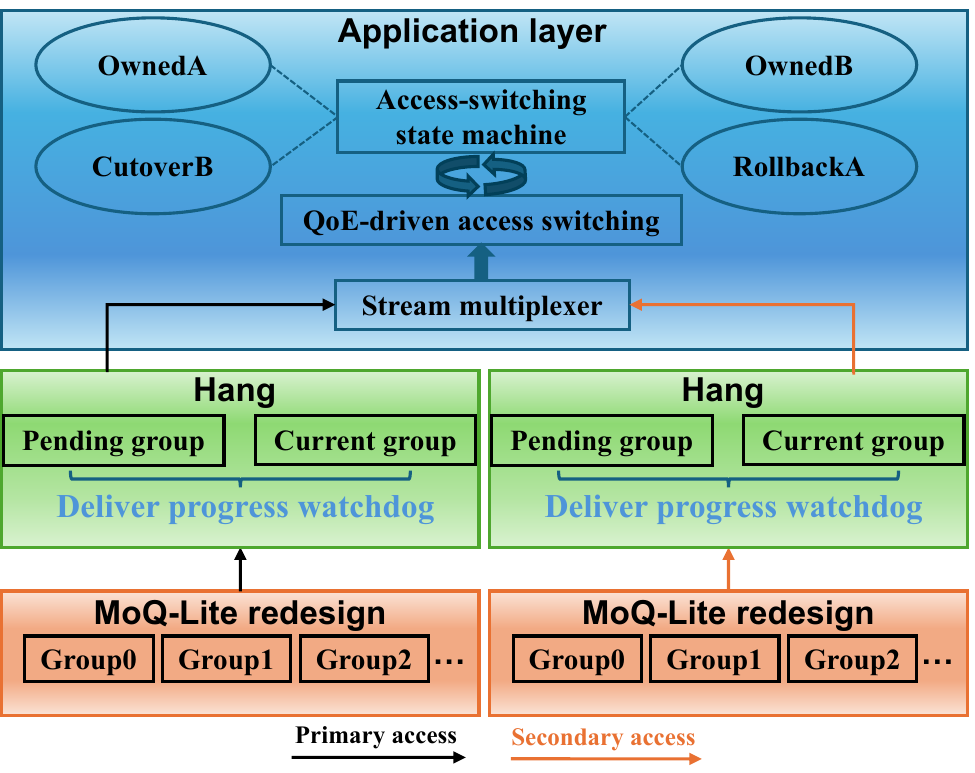}
  \caption{Client-driven multi-access control.}
  \Description{Two parallel paths represent primary and secondary access. Each path feeds redesigned MoQ-Lite groups into a Hang instance with pending and current group states monitored by a delivery-progress watchdog. A stream multiplexer combines the two paths into one logical stream. Above it, quality-driven access switching controls a state machine with OwnedA, CutoverB, OwnedB, and RollbackA states.}
  \label{fig:client_multi_access}
\end{figure}

\paragraph{MoQ-Lite/Hang decoupling.}
To remove the cross-layer race identified in \secref{subsec:motivation}, Hang consumes the receive queue and maintains its own \textit{pending} and \textit{current group} states (Fig.~\ref{fig:client_multi_access}). It does not lose its current group merely because MoQ-Lite receives a newer one. A \textit{delivery progress watchdog} instead monitors useful progress. If a current group remains blocked beyond the watchdog interval, Hang drops it and advances to the next pending group. This separates low-level arrivals from the upper-layer decision to abandon a group.

\paragraph{Stream multiplexing.}
Each access has its own Hang instance, but \system{} exposes one logical output per media track. The \textit{stream multiplexer} applies make-before-break switching: the current access continues to deliver while the alternate subscription warms up. Cutover occurs only after the alternate stream reaches a decoder-safe entry point, such as a keyframe. The multiplexer enforces monotonic output and a \textit{timestamp lead guard}, which requires the new stream to advance beyond the last timestamp committed from the old stream. It briefly drains the old access after cutover before unsubscribing. Duplicate or late objects are not exposed, so the application observes a continuous track.

\paragraph{QoE-driven switching.}
Each client runs a closed-loop controller per track. During startup, it records a latency baseline; during playback, it maintains a smoothed short-window latency estimate. A handover requires several signals to agree: relative and absolute increases over the baseline, a sustained upward trend, and confirmation in consecutive windows. This conservative trigger rejects transient jitter but responds when one access persistently degrades. A per-round limit on successful handovers also prevents many tile tracks from switching simultaneously.

\paragraph{Per-track state machine.}
A track starts in \textit{OwnedA}, with media delivered through the primary access. Persistent degradation moves it to \textit{WarmupB}, which subscribes to and probes the secondary while retaining the primary. Once the alternate stream satisfies the lead and decoder-entry conditions, \textit{CutoverB} commits its first accepted object and transitions to \textit{OwnedB}. A timeout, error, or insufficient progress invokes \textit{RollbackA} and retains the primary output. The application does not observe the failed attempt.

\noindent\textbf{\textit{Design effect.}}
The delivery redesign and endpoint controller decouple path selection from media continuity. The controller can evaluate and change accesses while the application retains one ordered stream. Multi-access operation also need not add proportional backhaul load under MoQ's relay-assisted fan-out. A secondary relay may already carry a requested tile for other clients; a new subscription can then reuse that relay distribution rather than starting another end-to-end retrieval. When the tile is absent, the additional subscription still follows standard MoQ relay behavior.

The controller can also retain single-access delivery when the primary path remains adequate. Multi-access is activated per track during warmup or handover rather than imposed continuously on the entire session. This keeps the access decision local to the affected tile while the multiplexer preserves the session-wide media interface.

\subsection{Implementation}
\label{subsec:implementation}

We implement \system{}\footnote{Source code: \url{https://github.com/YitongLI2000/MultiMoQ-ACMMM2026.git}.} as a layered extension of the Rust MoQ stack\footnote{\url{https://github.com/moq-dev/moq.git}.}, which follows the IETF MoQ draft~\cite{moq}. The prototype preserves the publisher--relay--client structure, publish/subscribe model, and QUIC transport. Pending/active scheduling and receive queues remain inside MoQ-Lite. Watchdogs, per-access Hang instances, the multiplexer, and the state machine run at client endpoints. Relays remain transport-facing forwarding nodes.

Thus, \system{} is an in-stack evolution of MoQ rather than a replacement transport. Existing publisher and relay roles remain intact, and the new access logic is confined to endpoints.
The implementation therefore changes delivery control without requiring a new wire transport or a media-aware relay API.

\section{Evaluation}
\label{sec:evaluation}

We evaluate three questions. First, does \system{} improve goodput, latency, and continuity across enhancement, base, and audio tracks? Second, which gains come from the MoQ-Lite redesign and which require client-driven multi-access control? Third, does \system{} improve bottleneck delivery and latency jointly, and do these gains survive viewport reconstruction?

\subsection{Experiment Setup}
\label{subsec:experiment_setup}

\subsubsection{Media Sample}
\label{subsubsec:media}

We used a 4K, 30-fps 360\textdegree{} ERP video. FFmpeg\footnote{\url{https://www.ffmpeg.org/}} partitioned each frame into a $4\times2$ grid and packaged eight enhancement tiles, one base track, and one audio track as fMP4. The $2048\times1024$ base track was encoded at 6.08~Mbps, while each $1024\times1024$ enhancement tile was encoded at 2.07--2.46~Mbps. Audio used 48-kHz AAC-LC at 0.19~Mbps. Video totaled 24.44~Mbps. Keyframes occurred at roughly 1-s intervals, and we aligned both DASH-segment and MoQ-group boundaries with these intervals so that the baselines used the same media units and random-access points.

\subsubsection{Evaluation Environment and Streaming Setup}
\label{subsubsec:streaming_setup}

Experiments ran in Mininet~\cite{mininet} on a 128-core Intel Xeon server with 256~GB RAM. \system{} and standard MoQ used identical packaging, with one 1-s group of pictures (GOP) per MoQ group. The DASH baseline used tiled HTTP/2 over TCP~\cite{http2}, a dynamic Media Presentation Description, and 1-s segments aligned with the GOPs\footnote{\url{https://dashif.org/dash.js/pages/usage/low-latency.html}}. The base track, enhancement tiles, and audio appeared as independent tracks in both systems.

Each DASH client first fetched the initialization segment from its assigned CDN proxy and then requested the latest unseen media segment for every track. We disabled adaptive bitrate switching because the study isolates transport and delivery behavior rather than representation selection. Nginx reverse-proxy caches provided a shared 30-MB budget per proxy across concurrent clients\footnote{Implementation based on \url{https://github.com/TareqAlqutami/rtmp-hls-server}.}. This constrained cache allowed concurrent tile requests to compete for the same edge capacity.

\subsubsection{Baselines and Evaluation Metrics}
\label{subsubsec:baseline_metrics}

We compared \textbf{DASH}, standard \textbf{MoQ}, the \textbf{MoQ-Lite Redesign-only} variant, and the full \textbf{\system{}}. The redesign-only ablation includes the pending/active and receive queues but omits client-driven multi-access control, isolating the contributions of stable group delivery and access adaptation.

We evaluate performance using four metrics:
\begin{itemize}[leftmargin=1.2em,itemsep=1pt,topsep=2pt,parsep=0pt,partopsep=0pt]
    \item \textbf{Transport stall time}: time that delivery blocks because insufficient media has arrived.
    \item \textbf{Goodput}: useful media data delivered to the client per unit time.
    \item \textbf{End-to-end latency}: elapsed time from source generation to client delivery.
    \item \textbf{Viewport quality}: freeze behavior and the SSIM/VMAF fidelity of reconstructed viewports against the source.
\end{itemize}

\subsubsection{Network Topology}
\label{subsubsec:topology}

\begin{figure}[ht]
  \centering
  \includegraphics[width=1\linewidth]{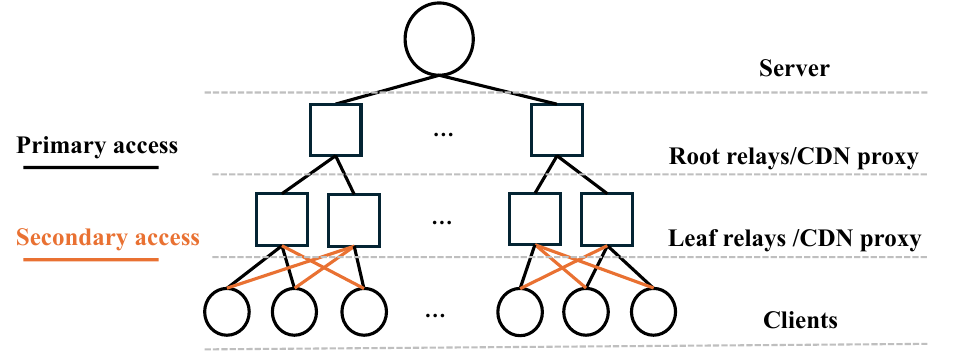}
  \caption{Common four-layer topology for the MoQ and DASH evaluations.}
  \Description{A four-layer tree consisting of one server, root relays or CDN proxies, leaf relays or CDN proxies, and clients. Black links form the hierarchical primary-access paths from each client through a leaf and root node to the server. Orange links cross-connect clients to alternate leaf nodes to provide secondary access. Ellipses indicate additional nodes at each layer.}
  \label{fig:eval_topology}
\end{figure}

Fig.~\ref{fig:eval_topology} shows the common four-layer topology: one server, four root nodes, 16 leaf nodes, and 96 clients. Each root connected to four leaves; each leaf was the primary node for six clients and the secondary node for six others. MoQ used relays and DASH used CDN proxies at the same positions, so both baselines traversed the same hierarchy and fan-out.

Primary-access links followed the measured bandwidth and delay distributions in Table~\ref{tab:campus_network}. Fig.~\ref{fig:eval_viewport_quality}(a) shows the resulting per-leaf bandwidth assignment. Every secondary link used 30~Mbps, with delay drawn from the table's Tencent-CDN range. All upper-layer backhaul links used 80~Mbps and 2~ms delay to represent a constrained shared delivery network~\cite{topo-backhaul}. Unless stated otherwise, every experiment used this topology.

\subsection{Overall Transport Performance}
\label{subsec:eval_transport_performance}

\begin{figure*}[!htbp]
    \begin{minipage}[t]{\textwidth}
        \centering
        \includegraphics[width=\textwidth]{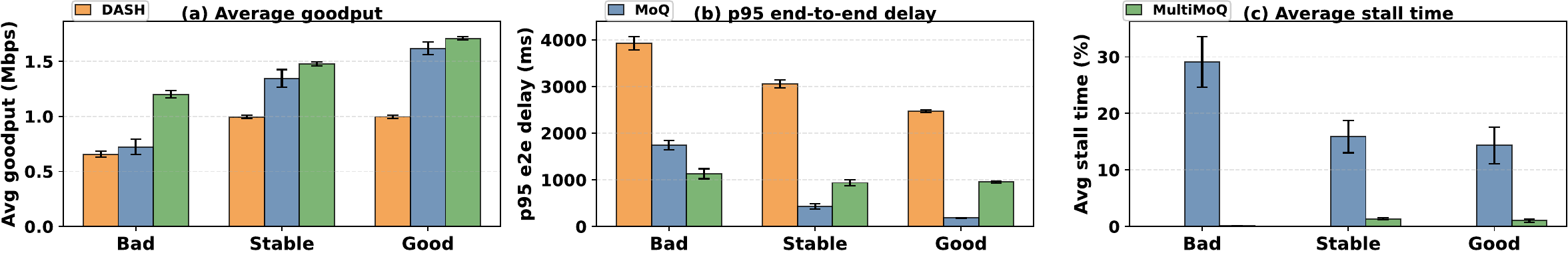}
        \caption{\system{} improves enhancement-tile goodput and tail delay while keeping stalls near zero.}
        \Description{Three grouped bar charts compare DASH, standard MoQ, and MultiMoQ under bad, stable, and good networks, with error bars. MultiMoQ has the highest average enhancement-tile goodput in all three conditions. Its 95th-percentile end-to-end delay is much lower than DASH, although standard MoQ is lower under stable and good conditions. Standard MoQ has average stall time between roughly 14 and 29 percent, whereas DASH and MultiMoQ remain close to zero.}
        \label{fig:eval_overall_conclusion}
    \end{minipage}
\end{figure*}

\begin{figure*}[!htbp]
    \begin{minipage}[t]{\textwidth}
        \centering
        \includegraphics[width=\textwidth]{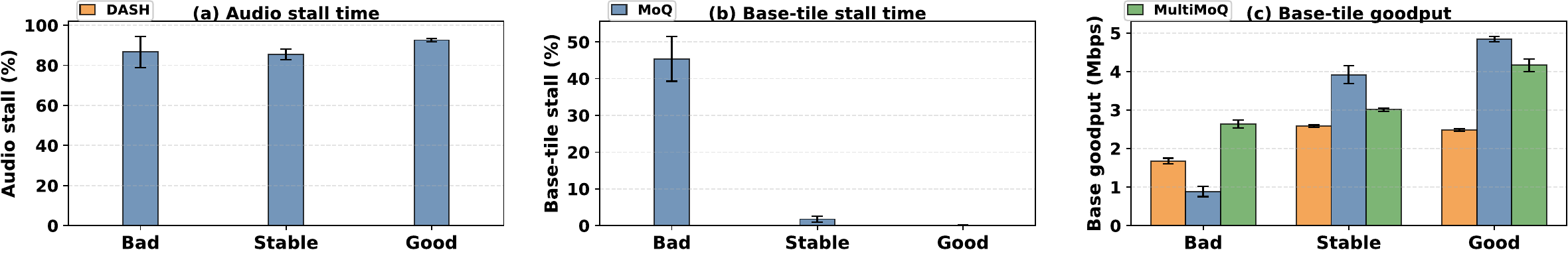}
        \caption{\system{} preserves audio and base-track continuity across network conditions.}
        \Description{Three grouped bar charts compare DASH, standard MoQ, and MultiMoQ. Standard MoQ has audio stall time above 85 percent in every network condition, while DASH and MultiMoQ are near zero. Standard MoQ also has about 45 percent base-tile stall in the bad network and much less in better networks. MultiMoQ delivers higher base-tile goodput than DASH in all conditions, while standard MoQ is highest under stable and good conditions.}
        \label{fig:eval_audio_base}
    \end{minipage}
\end{figure*}

\begin{figure*}[t]
  \centering

  \begin{minipage}[t]{0.48\textwidth}
    \centering
    \includegraphics[width=\linewidth]{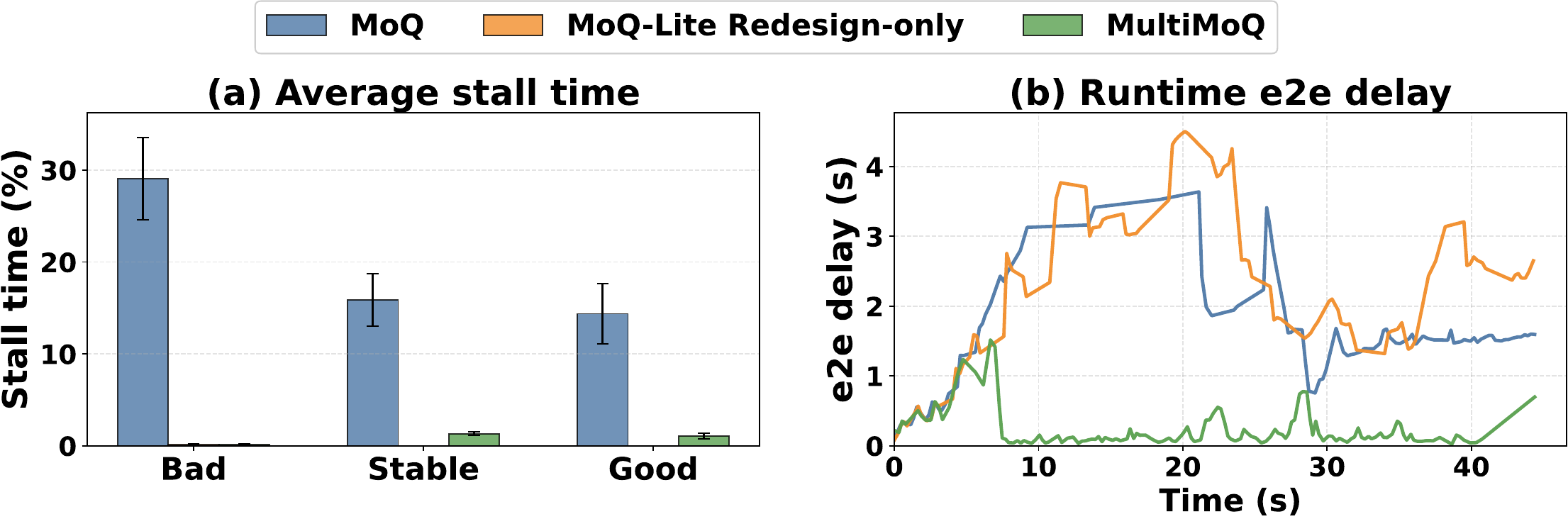}
    \captionof{figure}{Transport ablation: stall time and runtime delay.}
    \Description{Two plots compare standard MoQ, redesign-only, and MultiMoQ. Standard MoQ has average stall times of 29.1, 15.9, and 14.4 percent, while redesign-only and MultiMoQ remain near zero across bad, stable, and good access. The runtime trace shows delay rising for standard MoQ and redesign-only, whereas MultiMoQ's delay falls after access switching near 7.2 seconds.}
    \label{fig:eval_ablation_stall}
  \end{minipage}
  \hfill
  \begin{minipage}[t]{0.48\textwidth}
    \centering
    \includegraphics[width=\linewidth]{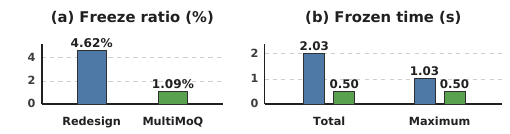}
    \captionof{figure}{Playback ablation: freeze ratio and duration.}
    \Description{Two bar charts compare redesign-only and MultiMoQ. Freeze ratio falls from 4.62 percent to 1.09 percent. Total frozen time falls from 2.03 seconds to 0.50 seconds, and maximum freeze duration falls from 1.03 seconds to 0.50 seconds.}
    \label{fig:eval_ablation_freeze}
  \end{minipage}
\end{figure*}

\noindent\textbf{\textit{Enhancement-Tile Performance.}} 
Fig.~\ref{fig:eval_overall_conclusion} compares the eight enhancement tiles under the \textit{bad}, \textit{stable}, and \textit{good} access conditions. In the bad condition, \system{} delivered 1.20~Mbps versus 0.66~Mbps for DASH and 0.72~Mbps for standard MoQ. Its goodput rose to 1.48 and 1.71~Mbps in the stable and good conditions, remaining above both baselines.

Relative to DASH, \system{} reduced p95 end-to-end delay from 3930/3062/2473~ms to 1130/936/956~ms across the three conditions. Standard MoQ had lower delay in the latter two conditions, but this result came with poor continuity: it stalled for 29.1\% in the bad condition and above 14\% otherwise. \system{}'s corresponding stall times were 0.11\%, 1.32\%, and 1.03\%. DASH also rarely stalled, but delivered lower goodput at much higher tail delay. \system{} therefore retained MoQ's timely object delivery without its persistent stalls.

\noindent\textbf{\textit{Audio and Base Track Robustness.}} 
Fig.~\ref{fig:eval_audio_base} shows whether the same behavior holds for audio and fallback base video. Standard MoQ's audio stall time was above 85\% in every condition, whereas DASH and \system{} remained near zero. Its base track also stalled about 45\% in the bad condition; the other two systems again remained near zero.

In the bad condition, \system{} increased DASH base-track goodput from 1.69 to 2.65~Mbps. The corresponding gains were from 2.59 to 3.02~Mbps in the stable condition and from 2.49 to 4.16~Mbps in the good condition. Standard MoQ delivered more base-track goodput in the stable and good cases, but with poor audio continuity and severe bad-condition base-track stalls. The redesign thus benefited short audio groups and the base track in addition to enhancement tiles.

\subsection{Ablation Study}
\label{subsec:eval_ablation}

In Fig.~\ref{fig:eval_ablation_stall}(a), the redesign-only variant reduced standard MoQ's 29.1\%/15.9\%/14.4\% stall times to 0.15\%/0.00\%/0.00\%. The MoQ-Lite redesign therefore accounts for most of the continuity improvement. Eliminating stalls, however, did not guarantee timely playback. Fig.~\ref{fig:eval_ablation_stall}(b) shows that redesign-only delivery still accumulated several seconds of runtime delay on a degraded access. Full \system{} reduced the delay after switching to a better access at $t=7.2$~s. For a conservative comparison, the standard-MoQ trace uses its lowest-delay tile.

Both variants use the same group-delivery redesign, so the post-switch divergence isolates client-driven access control.

Fig.~\ref{fig:eval_ablation_freeze} shows the application-level effect. Multi-access control reduced the redesign-only freeze ratio from 4.62\% to 1.09\%, maximum freeze from 1.03~s to 0.50~s, and total frozen time from 2.03~s to 0.50~s. The traces cover 44~s and 46~s, respectively, so freeze ratio normalizes their durations. Together, Figs.~\ref{fig:eval_ablation_stall} and~\ref{fig:eval_ablation_freeze} show that the redesign removed transport stalls while switching curtailed delay-related freezes.

\subsection{Multi-Track Delivery and Playback Analysis}
\label{subsec:eval_multitrack_playback}

\begin{figure*}[t]
  \centering
  \includegraphics[width=\textwidth]{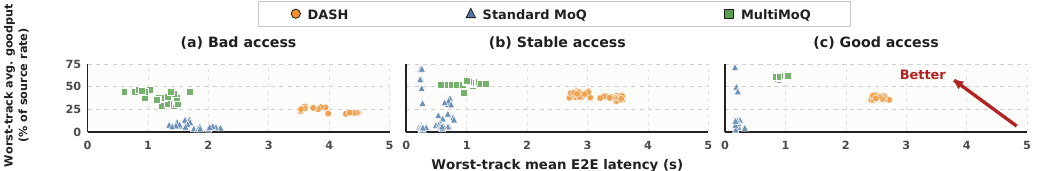}
  \caption{Per-client delivery across eight tracks: minimum source-normalized mean goodput versus maximum mean latency.}
  \Description{Three scatter plots compare DASH, standard MoQ, and MultiMoQ under bad, stable, and good access. For each client, the vertical coordinate is the lowest ratio of measured mean goodput to source encoding rate among video2 through video9, while the horizontal coordinate is the greatest mean end-to-end latency among those tracks. Better outcomes lie toward the upper left. DASH clusters at higher latency, standard MoQ usually has a severely underdelivered bottleneck track, and MultiMoQ combines stronger bottleneck delivery with substantially lower latency than DASH.}
  \label{fig:eval_bottleneck_delivery}
\end{figure*}

\begin{figure}[ht]
  \centering
  \includegraphics[width=1\linewidth]{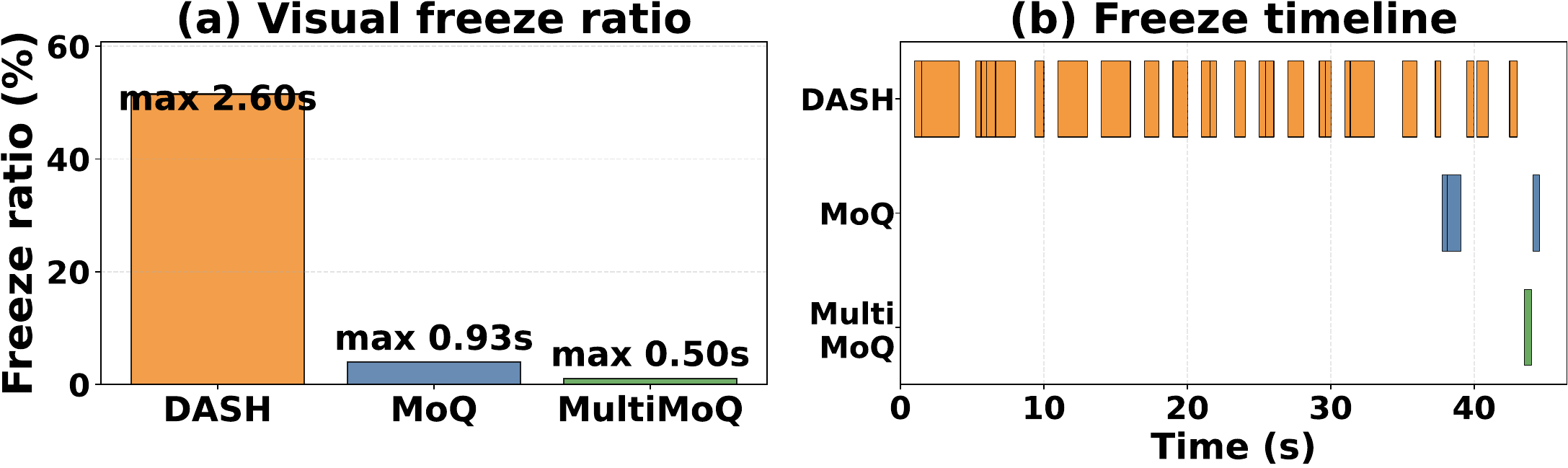}
  \caption{\system{} has the lowest freeze ratio and shortest maximum freeze.}
  \Description{A bar chart and timeline compare playback freezes for DASH, standard MoQ, and MultiMoQ over approximately 45 seconds. DASH has a 51.5 percent freeze ratio and a 2.60-second maximum freeze, with frequent freezes throughout the timeline. Standard MoQ has a 3.96 percent ratio and a 0.93-second maximum, with a few late freezes. MultiMoQ has a 1.09 percent ratio and one 0.50-second freeze near the end.}
  \label{fig:eval_playback_analysis}
\end{figure}

\begin{figure*}[!t]
  \centering
  \includegraphics[width=\textwidth]{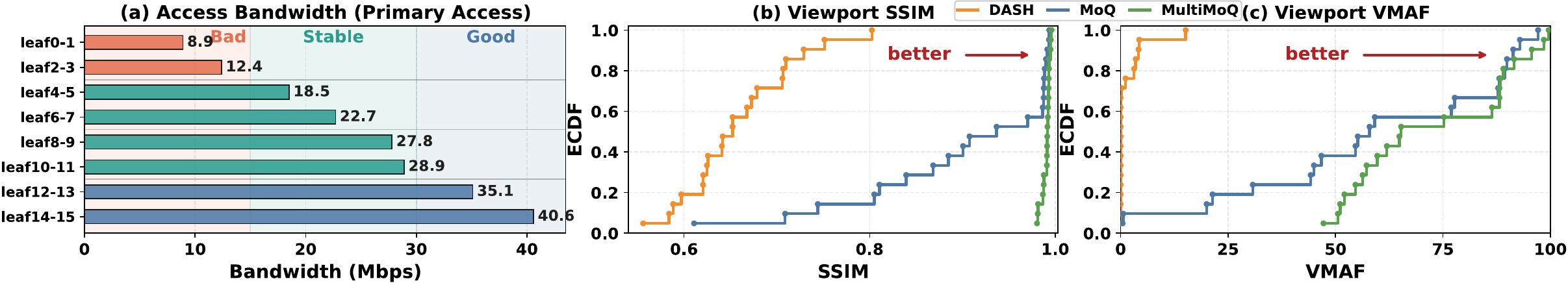}
  \caption{Access assignment and quality for 21 viewports: \system{} shifts SSIM and VMAF toward higher scores.}
  \Description{Three panels summarize the access configuration and viewport fidelity. The left panel assigns primary-access bandwidths of 8.9, 12.4, 18.5, 22.7, 27.8, 28.9, 35.1, and 40.6 Mbps to eight leaf pairs spanning the bad, stable, and good regimes. The middle and right panels show empirical cumulative distributions of viewport SSIM and VMAF for DASH, standard MoQ, and MultiMoQ. MultiMoQ is concentrated near the highest SSIM values and avoids the low-VMAF tail of the baselines.}
  \label{fig:eval_viewport_quality}
\end{figure*}

Fig.~\ref{fig:eval_bottleneck_delivery} reports 96 client observations per system from one shared-topology run, not independent repetitions. Let $\mathcal{T}$ contain video2--video9. The vertical coordinate is $B_i=100\min_{t\in\mathcal{T}}\bar{G}_{i,t}/R_t$; the horizontal coordinate is $D_i=\max_{t\in\mathcal{T}}\bar{L}_{i,t}$. Here $\bar{G}_{i,t}$ and $\bar{L}_{i,t}$ are track $t$'s mean goodput and latency, and $R_t$ is its source encoding rate. These rates span 2.07--2.46~Mbps; normalization makes differently encoded tiles comparable. Thus, $B_i$ is the weakest track's delivery percentage and $D_i$ its worst mean latency. Base video and audio are excluded, and the extrema can come from different tracks.

From bad through stable to good access, \system{}'s median $B_i$ rises through 39.4/52.5/60.5\% while $D_i$ stays near 1~s. DASH provides moderate bottleneck delivery (22.4--39.1\%) but 2.54--3.96~s latency; standard MoQ reaches 0.19--1.73~s but leaves $B_i$ at only 6.0--12.8\%. DASH therefore trades freshness for coverage and standard MoQ does the reverse. \system{} shifts outcomes upward relative to both and leftward relative to DASH, giving the strongest joint balance.

GStreamer\footnote{\url{https://gstreamer.freedesktop.org/}} decoded and composed the captured tiles into re-encoded ERP video. Holding the source, client, and stable access condition fixed, we evaluated 21 matched viewports at seven timestamps and three yaw angles ($-90^\circ$, $0^\circ$, and $90^\circ$). Each $1280\times720$ viewport used a $100^\circ\times100^\circ$ field of view; we computed frame-level SSIM and 1-s clip-level VMAF against the source.

Fig.~\ref{fig:eval_playback_analysis} quantifies continuity. DASH had a 51.5\% freeze ratio, 26 events, a 2.60-s maximum, and 22.67~s total frozen time. Standard MoQ had a 3.96\% ratio, three events, a 0.93-s maximum, and 1.77~s total. \system{} had one 0.50-s freeze and a 1.09\% ratio. The transport-level continuity and latency improvements therefore remained visible after tile decoding and viewport composition.

Fig.~\ref{fig:eval_viewport_quality}(b)--(c) quantifies this comparison. \system{}'s SSIM values were concentrated near 1.0, with a mean of 0.990 and a minimum of 0.980. Its VMAF distribution shifted toward higher scores than both baselines, avoiding DASH's low-score tail. The perceptual metrics therefore confirm that the transport gains preserve viewport fidelity in addition to playback continuity.

Fig.~\ref{fig:eval_bottleneck_delivery} treats all eight enhancement tiles symmetrically and is neither a main-tile nor a viewport-weighted metric. As a receiver-side aggregate, it omits compositor-selected timestamps, same-timestamp coverage, base fallback, missing viewport area, and inter-tile synchronization; Fig.~\ref{fig:eval_viewport_quality} separately measures sampled viewport fidelity.

\subsection{Discussion}
\label{subsec:eval_discussion}

The results expose two bottlenecks. The MoQ-Lite redesign stabilized concurrent tracks by preserving active groups, but redesign-only delivery still accumulated delay on a degraded access. The controller explains the runtime-delay and freeze reductions, while full \system{} also strengthened the weakest enhancement track.

The baselines expose the corresponding tradeoff: DASH avoided most transport stalls by buffering and falling behind, whereas standard MoQ remained timely for received objects but lost concurrent-track continuity. Robust immersive playback required both continuous group delivery and timely access adaptation; either mechanism alone left one failure mode unresolved.

Our scope is bounded to Mininet, one 4K source, one hierarchical topology, measured access distributions, and fixed 1-s media units, with bitrate adaptation disabled. The results isolate transport and delivery under these conditions rather than every Internet deployment; representation selection remains complementary to \system{}.

\section{Related Work}
\label{sec:related_work}

\paragraph{Viewport-adaptive streaming.}
Viewport-adaptive systems partition a panorama into tiles, predict the viewer's orientation, and vary quality spatially or temporally to reduce bandwidth~\cite{taxonomy-applayer1,taxonomy-applayer2,taxonomy-applayer3,taxonomy-applayer4,taxonomy-applayer5}. OMAF- and CMAF-based systems extend this approach to standards-compliant low-latency streaming~\cite{taxonomy-applayer-omaf1,taxonomy-applayer-omaf2,taxonomy-applayer-cmaf1}. These techniques decide which media to deliver, whereas \system{} changes how concurrent selected tiles are transported. Most prior systems still request segments or chunks over DASH-style HTTP, whose delivery granularity limits responsiveness to network variation.

These selection techniques are complementary to \system{}. A viewport predictor or representation controller can choose the tile set before \system{} schedules the resulting tracks across accesses. Their prediction and quality-allocation policies remain unchanged by our transport design.

\paragraph{QUIC-based transport.}
QUIC and multipath-QUIC systems apply cross-layer scheduling and packet or path selection under dynamic conditions~\cite{taxonomy-transport1,taxonomy-transport2,taxonomy-transport3}. Other systems run DASH or end-to-end 360\textdegree{} pipelines over QUIC and HTTP/3~\cite{taxonomy-transport4,taxonomy-transport5}. This work improves transport efficiency and path use, but generally retains a direct client--server architecture. \system{} instead operates within MoQ's relay tree and coordinates path changes with per-track media continuity.

The distinction is architectural: multipath transport schedules packets within an end-to-end connection, while \system{} switches application-visible media subscriptions across relay-assisted accesses. It also maintains decoder-safe, monotonic output across a switch rather than exposing independent path streams to the player.

\paragraph{Network-assisted delivery.}
Multi-source and distributed-node systems divide tiles across sources~\cite{taxonomy-other1,taxonomy-other2}. Multicast and collaborative systems share delivery across users~\cite{taxonomy-other3,taxonomy-other4}, while edge-assisted systems cache adaptive or live 360\textdegree{} content~\cite{taxonomy-other-edge1,taxonomy-other-edge2,taxonomy-other-edge3}. These architectures reduce pressure on one origin but require coordination at servers, edges, or the network. \system{} retains MoQ's relay-assisted fan-out, changes group delivery inside MoQ-Lite, and places multi-access decisions at the client. It does not require multicast or a new edge-coordination protocol.

Consequently, network-assisted delivery and \system{} address different control points. The former decides where content is placed or shared; the latter preserves group continuity and selects among accesses after relay distribution is available.

\section{Conclusion}
\label{sec:conclusion}

In this paper, we present \system{}, a MoQ-based framework for robust immersive video streaming under heterogeneous network conditions. By combining more stable MoQ-based delivery with client-driven multi-access adaptation, \system{} enables live high-bitrate 360\textdegree{} streaming to remain both responsive and robust under varying access quality. Our implementation and evaluation show that this design improves end-to-end streaming performance and bottleneck delivery across concurrent enhancement tracks, translating into smoother playback and higher viewport fidelity than DASH and standard MoQ. The ablation further confirms the contribution of multi-access control to playback continuity. These results suggest that MoQ, when extended with stronger delivery stability and access adaptation, provides a promising foundation for scalable, high-quality live immersive streaming.

\clearpage

\bibliographystyle{ACM-Reference-Format}
\balance
\bibliography{reference}

\end{document}